%% file: main.tex
\documentclass[screen, nonacm, sigconf]{acmart}

\usepackage{setspace}
\usepackage[shortlabels]{enumitem}
\usepackage[skins,breakable]{tcolorbox}
\usepackage{booktabs}
\usepackage{multicol}
\usepackage{multirow}
\usepackage[capitalize]{cleveref}
\usepackage{graphicx}
\usepackage{caption}

\usepackage{amssymb}

\usepackage{amsthm} 
\usepackage{amsmath,amsfonts}
\usepackage{pifont}
\usepackage{fontawesome5}
\usepackage{balance}

\usepackage[shortlabels]{enumitem}
\setenumerate[1]{itemsep=0pt,partopsep=0pt,parsep=\parskip,topsep=0.5pt}
\setitemize[1]{itemsep=0pt,partopsep=0pt,parsep=\parskip,topsep=0.5pt}
\setdescription{itemsep=0pt,partopsep=0pt,parsep=\parskip,topsep=0.5pt}

\newtheorem{example}{\bf Example}

\newcommand\blfootnote[1]{%
  \begingroup
  \renewcommand\thefootnote{}\footnote{#1}%
  \addtocounter{footnote}{-1}%
  \endgroup
}

\newcommand{\task}{\textsc{nlcTD}}
\newcommand{\method}{\textsc{Crofuma}}
\newcommand{\demo}{\textsc{TableCopilot}}
\newcommand{\dataset}{\textsc{nlcTables}}

\newcommand{\remarkbox}[1]{
\vspace{-\topsep}
\begin{tcolorbox}[colback=gray!10, colframe=gray!50, coltitle=black, width=\columnwidth, boxrule=0.5pt, arc=1mm, boxsep=0mm, left=1.5mm, right=1.5mm, top=1.5mm, bottom=1.5mm, breakable]
{#1}
\end{tcolorbox}
\vspace{-\topsep}
}

\begin{document}
\title{\demo{}: A Table Assistant Empowered by Natural Language Conditional Table Discovery}







\author{Lingxi Cui}
\affiliation{%
  \institution{Zhejiang University}
  \city{Hangzhou}
  \country{China}
}
\email{cuilingxi.cs@zju.edu.cn}

\author{Guanyu Jiang}
\affiliation{%
  \institution{Zhejiang University}
  \city{Hangzhou}
  \country{China}
}
\email{django040805@zju.edu.cn}

\author{Huan Li}
\authornote{Huan Li and Lidan Shou are the corresponding authors.}
\affiliation{%
  \institution{Zhejiang University}
  \city{Hangzhou}
  \country{China}
}
\email{lihuan.cs@zju.edu.cn}

\author{Ke Chen}
\affiliation{%
  \institution{Zhejiang University}
  \city{Hangzhou}
  \country{China}
}
\email{chenk@zju.edu.cn}

\author{Lidan Shou}
\authornotemark[1]
\affiliation{%
  \institution{Zhejiang University}
  \city{Hangzhou}
  \country{China}
}
\email{should@zju.edu.cn}

\author{Gang Chen}
\affiliation{%
  \institution{Zhejiang University}
  \city{Hangzhou}
  \country{China}
}
\email{cg@zju.edu.cn}

\input{content/0-abstract}

\maketitle

\input{content/1-intro}
\input{content/2-sys}

\input{content/3-demo}
\input{content/4-Conclusion}

\begin{acks}
 This work was supported by the Major Research Program of the Zhejiang Provincial Natural Science Foundation (Grant No.~LD24F020015), the Pioneer R\&D Program of Zhejiang (Grant No.~2024C01021), NSFC Grant No.~U24A201401, and Zhejiang Province ``Leading Talent of Technological Innovation Program'' (No. 2023R5214).
\end{acks}

\begingroup
\bibliographystyle{ACM-Reference-Format}
\bibliography{main}
\endgroup

\balance

\end{document}

%% file: content/0-abstract.tex
\begin{abstract}

    The rise of LLM has enabled natural language-based table assistants, but existing systems assume users already have a well-formed table, neglecting the challenge of table discovery in large-scale table pools. To address this, we introduce \demo{}, an LLM-powered assistant for interactive, precise, and personalized table discovery and analysis.
    We define a novel scenario, \task{}, where users provide both a natural language condition and a query table, enabling intuitive and flexible table discovery for users of all expertise levels. To handle this, we propose \method{}, a cross-fusion-based approach that learns and aggregates single-modal and cross-modal matching scores. Experimental results show \method{} outperforms SOTA single-input methods by at least 12\% on NDCG@5.
    We also release an instructional video, codebase, datasets, and other resources on GitHub to encourage community contributions. \demo{} sets a new standard for interactive table assistants, making advanced table discovery accessible and integrated.\blfootnote{The definitive Version
of Record was published in VLDB'25, \url{https://dl.acm.org/doi/10.1145/xxxx}.}

\end{abstract}



%% file: content/1-intro.tex
\section{Introduction}
\label{sec:intro}


The rise of large language models (LLMs) has extended their capabilities to tabular data, enabling the development of table assistants for tasks such as TableQA~\cite{zhu2024autotqa}, tabular data manipulation~\cite{TableLLM2024Zhang} and analytics~\cite{wangidatalake}. 
These approaches have achieved notable success in downstream tasks, often assuming that users already have a table with sufficient data to work with. However, in many real-world scenarios, users may start without a table or need to enhance an existing one before the analysis begins~\cite{Cui2024TabularDA}. This introduces the need for an additional, often overlooked step: \textbf{table discovery} over large-scale table pools, which has received limited attention in the context of table assistants.


To empower table assistants with table discovery capabilities, we face two primary challenges.
(1) \textbf{\textit{LLM Incompatibility with Table Discovery}}: The foundation of table assistants, LLMs, are not inherently optimized for table discovery tasks. Importing a large-scale pool with millions of tables into an LLM is impractical due to token limits and privacy concerns. Moreover, LLMs trained primarily on sequential textual data struggle to interpret structured, tabular data, which are of two-dimensional nature. This necessitates the use of specialized table discovery methods. 
(2) \textbf{\textit{Gaps in Existing Table Discovery Methods}}: Current methods typically rely on either keyword(s) or a table as a query to retrieve relevant tables. However, keywords and single-table queries often fail to fully capture user intent, particularly in interactive table assistants where users may specify detailed natural language requirements (see Example~\ref{example:app}). Traditional methods focusing solely on keywords or query tables are therefore insufficient for such use cases.

\begin{example}\label{example:app} %
    Suppose a recruiter wants to populate a shortlist table with applicants from an application database for subsequent competence evaluation. This shortlist table includes fields such as ID, name, graduation year, CS grade, math grade, and phone number. The recruiter is looking for records that can be directly filled into this table, with a specific request: applicants must graduate in 2023 or 2024, and have a CS grade above 90. In this context, keyword-based table discovery methods cannot handle such complex and personalized NL requirements, and the retrieved tables cannot be directly unioned with the shortlist. Moreover, query-table-based discovery methods are even less applicable, as the retrieved tables cannot meet the NL requirements. Therefore, a table assistant with discovery capabilities is the ideal solution. It is designed for users unfamiliar with database structures, meets complex and personalized needs, and facilitates further analysis and processing of data.
\end{example}

To empower LLM-based table assistants with table discovery capabilities, we make the following efforts:

\noindent(1) \textbf{New User Scenario.}
We introduce a new user scenario: \emph{Natural Language Conditional Table Discovery} (\task{}), allowing users to provide both natural language (NL) requests and a query table.
Based on practical application needs, we define three key user cases within \task{}: \textbf{\emph{NL-only search}} that takes only NL as input; \textbf{\emph{NL-conditional table union search}} that targets row-based table unions, and \textbf{\emph{NL-conditional table join search}} that focuses on column-based table joins.
By incorporating NL conditions into \task{}, we bridge the gap between LLM-based table assistants and table discovery. This integration enables interactive, precise, and personalized searches, enhancing user experience.
Moreover, detailed NL conditions facilitate highly efficient searches across vast table repositories, reducing LLM calls and improving response speed, scalability, and cost-effectiveness.

\noindent(2) \textbf{Cross-Fusion-Based Solution.}
To address the challenges of this new practical scenario, where existing table discovery methods fail to simultaneously process query tables and NL conditions, we propose \method{}, a novel \emph{cross-fusion-based matching} method for table discovery.
\method{} calculates single-modal table matching scores and cross-modal NL condition matching scores separately, then aggregates them during online query processing (Section~\ref{subsec:sys_online}). 
\method{} can handle situations with only NL or a single table as input, making it compatible with traditional table discovery scenarios.
In the offline phase, we employ pretrained language models (PLMs) as encoders for both NL and tables. To better capture table structure and semantics, we enhance PLMs through contrastive learning during pretraining.
Table contents and metadata are encoded separately and indexed using HNSW~\cite{fan_semantics-aware_2023,dong_deepjoin_2023}, achieving an average query time of under 500 ms and enabling real-time interactions with LLMs (Section~\ref{subsec:sys_offline}).
Experimental results show that our \method{} outperforms SOTA single-input discovery methods by at least 12\% on NDCG@5 in the \task{} scenario (Section~\ref{ssec:exp}).

\noindent(3) \textbf{End-to-end Prototype.} 
We present \demo{}, a fully functional prototype that seamlessly integrates \method{} into an LLM-powered table assistant. \demo{} supports the entire workflow from NL-conditional table discovery to downstream applications like TableQA.
To ensure robustness and facilitate evaluation, \demo{} also incorporates existing table discovery methods for direct comparison (search panel in Figure~\ref{fig-demo}(a)).
This design demonstrates the practicality and superiority of equipping LLM-based assistants with native, efficient table discovery capabilities.


\smallskip
Our proposals offer significant technical advancements in two key areas:
\textbf{(a) LLM-based Table Assistants}. 
Recent studies have explored leveraging LLMs to enhance table-related tasks:
Zhu et al.~\cite{zhu2024autotqa} leverage multi-agent LLMs for TableQA.
\texttt{iDataLake}~\cite{wangidatalake} integrates LLMs into analytics systems for querying unstructured, semi-structured, and structured data in data lakes.
Zhang et al.~\cite{TableLLM2024Zhang} present \texttt{TableLLM}, an LLM-driven assistant capable of advanced operations like querying, updating, merging, and charting.
While these works excel in table analysis and manipulation, they can benefit from our solution, which introduces natural language conditional table discovery to enable seamless and comprehensive end-to-end table workflows.
\textbf{(b) Table Discovery}. 
Existing table discovery prototypes primarily focus on keyword-based or query-table-based search:
\texttt{Aurum}~\cite{castro_fernandez_aurum_2018} retrieves related tables by performing schema-based matching on user-provided keywords.
\texttt{Auctus}~\cite{castelo_auctus_2021} supports joinable and unionable table searches using query tables.
\texttt{LakeCompass}~\cite{lakecompass2024chai} combines keyword and query-table-based table search while considering follow-up ML task support.
However, these prototypes do not address the new \task{} scenario that allows searching tables with a query table conditioned by natural language.
Our proposal, with its tailored scheme, unlocks new possibilities for interactive, precise, and personalized table discovery.

%% file: content/2-sys.tex
\section{\mbox{\demo{}} and Key Techniques}
\label{sec:sys}

\subsection{System Overview}
\label{subsec:sys_overview}

\begin{figure}[]
  \includegraphics[width=\columnwidth]{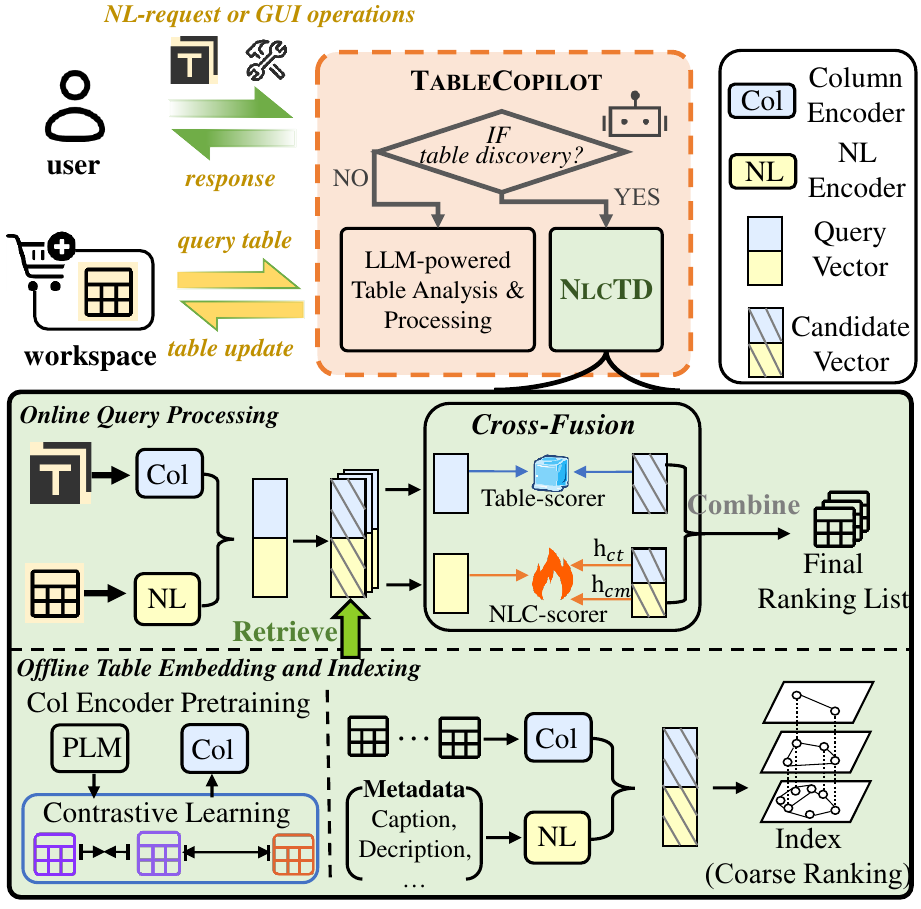}
  \caption{Overall Architecture of \demo{}.}
  \Description{}
  \label{fig-architecture}
\end{figure}

At a high level, \demo{} functions as an LLM-based table assistant capable of invoking table discovery methods, as shown in Figure~\ref{fig-architecture}. Users submit a natural language request or perform GUI actions, invoking the table assistant to respond and potentially update the workspace based on the user’s request and the current workspace status  (e.g., an empty or incomplete table for search, and a complete table for analysis and manipulation).

\smallskip
\noindent\textbf{Table Assistant Pipeline}. 
The assistant performs two key tasks: (1) identifying whether the human instruction involves a table discovery request and (2) handling table analysis and processing tasks beyond discovery. 
For a table discovery request, the assistant redirects the NL condition and a potential query table at the workplace to the online processing module of \method{}.
We employ an LLM, such as GPT-4o, to power the pipeline due to its ability to understand and process tables while following human instructions effectively. With carefully designed prompts, the LLM can accurately distinguish table discovery requests for appropriate redirection. Additionally, users can opt for alternative LLMs, including those fine-tuned specifically for table-related tasks~\cite{li_table-gpt_2023}.

\smallskip
\noindent\textbf{Processing \task{}}. 
Given a table repository $\mathcal{T}$, and a user query $q \in Q$ consisting of a query table $T^q$ and an NL condition $C$, the \task{} task aims to retrieve from $\mathcal{T}$ a top-$k$ ranked list of tables $\mathcal{T}' = \{ T_i \}$ that are \emph{semantically} relevant to both $T^q$ and $C$, as determined by a relevance scoring function, $\rho(T^q, C, T_i)$. 
Our \method{} follows a two-stage approach: (1) \textbf{\emph{the offline phase}} (Section~\ref{subsec:sys_offline}) encodes and indexes the entire table repository $\mathcal{T}$, and (2) \textbf{\emph{the online phase}} (Section~\ref{subsec:sys_online}) encodes the query and computes the similarity between $q$ and candidate tables $T_i$ for top-$k$ search.



\subsection{Offline Data Preparation}
\label{subsec:sys_offline}

\noindent\textbf{Column Encoder Pretraining}.
We adopt a column-based representation, which applies to both union (aggregating column similarity scores) and join (targeting join key column) searches within \task{}. 
We adopt contrastive learning, as used in \texttt{Starmie}~\cite{fan_semantics-aware_2023}, to train a PLM-based column encoder.
This allows the model to better capture contextual information, improving performance in cases where semantics are similar, but values differ.
Notably, our system is compatible with other column representation methods~\cite{dong_deepjoin_2023}, and users can integrate alternative column encoders.

\smallskip
\noindent\textbf{Table Repository Embedding}.
To facilitate \task{} processing, \method{} encodes both table content and metadata for all tables in the repository during the offline phase.
For \textbf{\emph{table embedding}}, we use the pretrained column encoder to derive the column embeddings and then concatenate them.
%
For \textbf{\emph{metadata embedding}}, we employ PLMs such as BERT or RoBERTa to encode textual information, including table captions and table descriptions.
These embeddings are then used to match natural language conditions during the online phase.
Finally, we concatenate the table embedding and the metadata embedding for subsequent indexing.

\smallskip
\noindent\textbf{Index Construction}.
Constructing an index for coarse ranking is crucial when searching large table repositories, especially in our prototype, which contains both column and metadata embeddings.
We choose to implement the Hierarchical Navigable Small World (HNSW) in \method{} due to its proven efficiency and lower loss rates~\cite{fan_semantics-aware_2023,dong_deepjoin_2023}. This index achieves an average query time of less than 500 ms for each in a repository containing 7,500+ tables.



\subsection{Online Query Processing}
\label{subsec:sys_online}


We propose a cross-fusion-based matching approach, \method{}, with two key modules: a table scorer (for query-table matching) and an NLC scorer (for NL-condition matching). These modules compute scores that are aggregated to produce the final result. This design seamlessly supports scenarios where only one input (query table or NL condition) is available, covering all three types of \task{} cases (Section~\ref{sec:intro}).
Before matching, the query table $T^q$ and NL condition $C$ are embedded using a method similar to offline embedding (Section~\ref{subsec:sys_offline}), with column embeddings $\mathbf{a}^i$ for $T^q$ and the NL condition embedding $\mathbf{c}$. To enhance efficiency, a limited set of candidate tables is first retrieved from the offline-constructed HNSW index, followed by cross-fusion matching on these candidates.

\smallskip
\noindent\textbf{Table Scorer} matches the query table with candidate tables using a learning-free method that directly computes similarity.
This approach is chosen as it operates within the same modality, leverages the pretrained table encoder, and minimizes online query time. \method{} supports both union and join scenarios, accounting for the diversity of NL conditions.
%

\textbf{\emph{NL-conditional Table Join Search}}:
For join scenarios, the focus is on matching values in the key column. We use the designated key column embedding $\mathbf{a}^k$ as the table embedding $\mathbf{t}^q$ and derive the table joinability score ${\rho}^t$ using cosine similarity.

\textbf{\emph{NL-conditional Table Union Search}}:
For union scenarios requiring multiple overlapping columns, column similarity scores are aggregated.
Inspired by previous works~\cite{fan_semantics-aware_2023,lakecompass2024chai}, we construct a bipartite graph $G = \langle V^q, V^T, E \rangle$, where nodes $V^q$ and $V^T$ represent the column sets of the query and candidate tables, and the edges $E$ represent column similarity scores. 
The table unionability score ${\rho}^t$ is computed by finding the maximum bipartite matching in $G$.

\smallskip
\noindent\textbf{NLC Scorer} matches the NL condition with candidate tables using a learning-based method for improved cross-modal matching. 
The NLC score is derived from two modules: condition-table matching and condition-metadata matching.

\textbf{\emph{Condition-Table Matching}}: This module models the interaction between the NL condition and a candidate table.
We concatenate the condition vector $\mathbf{c}$ and candidate table vector $\mathbf{t}_i$ along with their element-wise subtraction and Hadamard product\footnote{This concatenation method is proven effective to model embedding interactions~\cite{wang_retrieving_2021}.}:
\begin{equation}
\mathbf{\hat{h}}_i = \operatorname{concat} \big( \mathbf{t}_i, \mathbf{c}, (\mathbf{t}_i - \mathbf{c}), (\mathbf{t_i} \circ \mathbf{c}) \big).
\end{equation}
A nonlinear transformation is applied to produce hidden vectors:
\begin{equation}
\mathbf{h}_i = \operatorname{Tanh}(\textbf{W}_1 \hat{\textbf{h}}_i + \textbf{b}_1).
\end{equation}
Finally, we compute the hidden vectors for all $|\mathcal{T}|$ candidate tables using a max-pooling operation, which extracts the most relevant content for the NL condition~\cite{wang_retrieving_2021}:
\begin{equation}
\mathbf{h}^\mathit{ct} = \operatorname{MaxPooling}(\mathbf{h}_1, \mathbf{h}_2, \ldots, \mathbf{h}_{|\mathcal{T}|}).
\end{equation}

\textbf{\emph{Condition-Metadata Matching}}: This module matches the NL condition with the table metadata (e.g., table captions), which often reflects the table's content~\cite{lakecompass2024chai}. 
Treating this as a text matching task, we compute $\textbf{h}^\mathit{cm}$ using RoBERTa as the backbone, though models like T5 can also be used.

\begin{figure*}[]
  \includegraphics[width=0.98\textwidth]{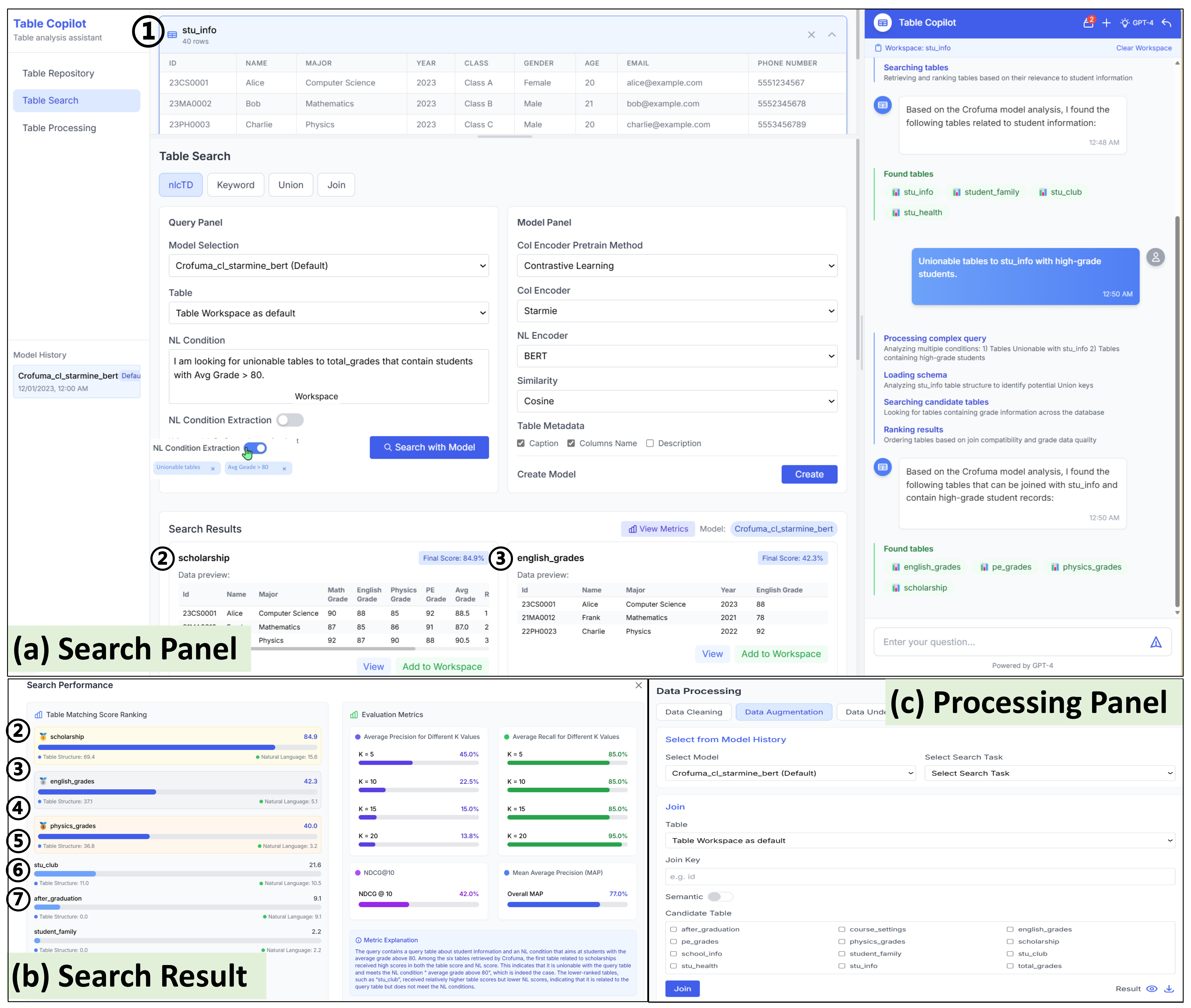}
  \caption{The Graphical User Interface of \demo{}. For a live demonstration, visit our project website on GitHub~\cite{demourl}.}
  \Description{}
  \label{fig-demo}
\end{figure*}

\smallskip
\noindent\textbf{Optimization of \method{}}.
The NLC score $\rho^c$ is obtained by concatenating $\textbf{h}_c = [\textbf{h}^\mathit{ct}; \textbf{h}^\mathit{cm}]$ and passing it through a multi-layer perceptron (MLP). 
The final score $\rho_k$ for each table $T_k$ is obtained by combining the table score $\rho^t$ and condition score $\rho^c$ with a weighting factor $\lambda$. This approach remains robust even if one of the scores is unavailable, ensuring \method{}'s compatibility with single-input table discovery.
To compare $\rho_k$ with the gold label $y_k$, we optimize the MLP using mean square error (MSE) loss, following previous table discovery studies~\cite{wang_retrieving_2021}:
\begin{equation}
\arg\min_{\theta} \frac{1}{|Q|} \sum\nolimits_{q \in Q} \frac{1}{|\mathcal{T}|} \sum_{k=1}^{|\mathcal{T}|} \big((\operatorname{MLP}(\textbf{h}_c, \theta) + \lambda \cdot \rho_{t}) - y_k\big)^2.
\end{equation}

\subsection{Effectiveness of \method{}}
\label{ssec:exp}

We evaluate our \method{} on the \dataset{} dataset~\cite{nlcTables}, which includes: \textbf{627} realistic queries covering NL-only, union, join, and fuzzy conditions; \textbf{22,080} tables from a large-scale repository with table sizes ranging from 1 to 69.5K rows and up to 33 columns; and \textbf{21,200 gold label annotations} with ample positive and negative ground truths for each query type.
We include six representative approaches published at top-tier venues, two each for keyword-based search (\texttt{GTR}~\cite{wang_retrieving_2021} and \texttt{StruBert}~\cite{trabelsi_strubert_2022}), table union search (\texttt{Santos}~\cite{khatiwada_santos_2023} and \texttt{Starmie}~\cite{fan_semantics-aware_2023}), and table join search (\texttt{Josie}~\cite{zhu_josie_2019} and \texttt{Deepjoin}~\cite{dong_deepjoin_2023}). 
As shown in Table~\ref{table:result}, our \method{} outperforms other single-input methods, demonstrating its superior ability to capture both query table and NL condition information, making it better suited for the new \task{} scenario.

\begin{table}[!htbp]
\caption{Comparisons of \method{} with SOTA methods.}
\label{table:result}
\resizebox{\columnwidth}{!}{%
\renewcommand{\arraystretch}{1}
\begin{tabular}{@{}lccllcc@{}}
\toprule
\multicolumn{3}{c}{nlc-Union} & \multirow{7}{*}{} & \multicolumn{3}{c}{nlc-Join} \\ \cmidrule(r){1-3} \cmidrule(l){5-7} 
          & NDCG@5   & NDCG@5 &                      &           & NDCG@5  & NDCG@5 \\
\texttt{Santos}    & 0.2076   & 0.2912 &                      & \texttt{Josie}     & 0.4691  & 0.5170  \\
\texttt{Starmie}   & 0.3066   & 0.4086 &                      & \texttt{Deepjoin}  & 0.2985  & 0.3959 \\
\texttt{GTR}       & 0.3211   & 0.4459 &                      & \texttt{GTR}       & 0.5071  & 0.5571 \\
\texttt{StruBert}  & 0.3675   & 0.3992 &                      & \texttt{StruBert}  &   0.5270      &   0.5291     \\
\textbf{\method{}}      & \textbf{0.4124}  & \textbf{0.4825} &                      & \textbf{\method{}}      & \textbf{0.6674}  & \textbf{0.6976} \\ \bottomrule
\end{tabular}
}
\end{table}




%% file: content/3-demo.tex
\section{Demonstration}
\label{sec:demo}

\noindent\textbf{Artifacts and Demonstration Materials}.
\demo{} is built on the \texttt{Flask} framework and comprises over \textbf{\emph{10,000 lines of backend and frontend code}}. To encourage participation from developers and researchers, we provide an instructional video, the codebase, and additional resources on GitHub~\cite{demourl} to foster exploration of advanced table assistants capable of table discovery.

\smallskip
\noindent\textbf{Runtime Case Study}.
Figure~\ref{fig-demo} demonstrates a real application of \method{} to the \task{} scenario, showcasing its ability to handle complex NL conditions.
The query involves a query table about student information (Table \ding{172} in Figure~\ref{fig-demo}(a)) and an NL condition that specifies: ``Find unionable tables containing students with an average grade above 80.''
\textbf{\emph{Search Panel}} (Figure~\ref{fig-demo}(a)): Users input the query table and define the NL condition via an intuitive interface. The search panel allows selection of algorithms, such as \method{}, with the model panel inside providing flexibility in configuring search parameters.
\textbf{\emph{Search Results}} (Figure~\ref{fig-demo}(b)): Among the six retrieved tables (Table \ding{173} - \ding{178} in Figure~\ref{fig-demo}(b)), the top-$1$ table ``scholarship'' (Table \ding{173}) received high scores for both table score and NL score, confirming it is unionable with the query table and satisfies the NL condition.
Lower-ranked tables, such as ``english\_grades'' (Table \ding{174}), scored high in table scores but failed to meet the NL condition, as reflected by their low NL scores.
\textbf{\emph{Processing Panel}} (Figure~\ref{fig-demo}(c)): After identifying relevant tables, users can proceed with further data processing tasks like cleaning and augmentation. This panel enables seamless integration of discovery and processing workflows, ensuring a streamlined user experience.

\smallskip
\noindent\textbf{Key Benefits for the Audience}.
\demo{} provides a distinctive platform to experience LLM-powered data management via an intuitive and engaging interface:
\begin{enumerate}[label=\textbf{\arabic*.}, leftmargin=1.5em]
    \item \textbf{Accessible and Flexible \task{} Scenario}.  
    Explore a highly user-friendly interface (Figure~\ref{fig-demo}(a)) that supports flexible NL condition edits on themes, categories, and distributions. The LLM enables natural language interactions, making advanced table discovery accessible to users of all expertise levels.

    \item \textbf{Deep Insights into \method{}}.  
    Interact directly with \method{} through a dynamic model panel, comparing its performance against leading algorithms like \texttt{Starmie}. Visualizations of intermediate results (Figure~\ref{fig-demo}(b)) demonstrate how \method{} effectively handles complex queries involving both union and join tasks, outperforming traditional approaches.

    \item \textbf{Integrated Table Discovery and Analysis.}  
    Learn about seamless integration of LLMs in table discovery workflows. \demo{} automatically detects discovery needs, invokes optimal algorithms, and facilitates advanced table manipulation (Figure~\ref{fig-demo}(c)), connecting AI advances with database research.
    
\end{enumerate}

%% file: content/4-conclusion.tex
\section{Conclusion}
\label{sec:conclusion}


In this paper, we make the first attempt to facilitate LLM-based table assistant with the capability of table discovery. To achieve this, we first define a new table discovery scenario \task{} that takes both the query table and the NL condition as input and presents its corresponding method \method{}. Finally, we present \demo{}, a fully functional prototype that supports the end-to-end workflow from NL-conditional table discovery to downstream table manipulation and analysis applications. \demo{} introduces a new paradigm for table discovery, aligning with the growing emergence of AI-powered tools for data science.